\documentclass[aps,prl,twocolumn,lengthcheck]{revtex4}
\usepackage{amsfonts, amssymb, amsmath,graphicx}
\usepackage{subfigure}
\usepackage{comment}

\begin{document}

\title{Striped states in weakly trapped ultracold Bose gases with Rashba spin-orbit coupling}

\author{Tomoki Ozawa}
\author{Gordon Baym}
\affiliation{
Department of Physics, University of Illinois at Urbana-Champaign, 1110 West Green Street, Urbana, Illinois 61801, USA
}%

\date{\today}

\def\del{\partial}
\def\p{\prime}
\def\simge{\mathrel{%
         \rlap{\raise 0.511ex \hbox{$>$}}{\lower 0.511ex \hbox{$\sim$}}}}
\def\simle{\mathrel{
         \rlap{\raise 0.511ex \hbox{$<$}}{\lower 0.511ex \hbox{$\sim$}}}}
\newcommand{\feynslash}[1]{{#1\kern-.5em /}}

\begin{abstract}
The striped state of ultracold bosons with Rashba spin-orbit coupling in a homogeneous infinite system has, as we show, a constant particle flow, which in a finite-size system would accumulate particles at the boundaries; it is thus not a physical steady state of the system.  We propose, as a variational ansatz, a condensate wave function for a weakly trapped system which behaves similarly to the striped state near the center, but does not have particle flow at the boundaries.
This state has a line of unquantized coreless vortices.
We show, by minimizing the total energy, that our modified striped state has lower energy than the conventional striped state and it is thus a physically appropriate starting point to analyze striped states in finite systems.
\end{abstract}

\maketitle

With the recent experimental realization of non-Abelian gauge fields in ultracold atomic gases ~\cite{Lin2011, Williams2012, Wang2012arXiv}, and in particular Rashba-Dresselhaus spin-orbit coupling ~\cite{Rashba1960,Dresselhaus1955}, the properties of the novel phases possible in these systems have come to the fore.
Ultracold bosons with Rashba-Dresselhaus spin-orbit coupling have been predicted to exhibit, depending on the details of the interparticle interaction, a plane-wave state, which is a Bose-Einstein condensate (BEC) of a single momentum state, or a striped state, which is a BEC made of particles with two opposite momenta \cite{Stanescu2008,Ho2011,Wang2010,Zhai2012,Ozawa2012a}.
At nonzero temperature, exotic phases, such as a nematic superfluid phase~\cite{Jian2011} and a BEC stabilized  by interactions, are predicted~\cite{Ozawa2012b, Barnett2012}.  In trapped systems, even more exotic phases, such as a half-vortex phase and a lattice phase, have been predicted~\cite{Wu2011,Sinha2011,Hu2012,Xu2012}.
When the trap potential energy is small compared to the kinetic and the interparticle interaction energies, numerical calculations predict that in two dimensions phases similar to the plane-wave and the striped phases in a homogeneous system may emerge~\cite{Sinha2011,Hu2012}.

In a homogeneous system, the striped phase has, as we show, alternating local particle current along the stripes, which accumulate particles at the boundaries, and thus it cannot be a physically realizable steady state in a finite system.
In this paper, we address the question of how such a striped state can be modified to embed it into a finite system.   Two possible remedies are surface currents at the ends of the stripes which recirculate the particle flow, or flow between the stripes in bulk.  We explore this latter solution in this paper, considering a system trapped by a weak harmonic potential in the direction (along y) of the stripes.  For simplicity, in this first step, we do not confine the system transverse to the stripes.  We propose a variational ansatz for a modified striped  
 condensate wavefunction which behaves as the striped state near the center, but does not have a net current flow at the boundary.  The wavefunction is Gaussian in the direction of the stripes, with a complex phase describing divergence-free particle flow.  
Fixing the spatial extent of the Gaussian by minimizing the energy, we show that the modified striped state has lower energy than a simple striped state, and thus is a physically acceptable candidate for the ground state.

{\em Local current in the striped phase.}
We consider bosons in two hyperfine states, labeled $\uparrow$ and $\downarrow$, with pure Rashba spin-orbit coupling, described by the Hamiltonian
\begin{align}
	\mathcal{H}
	&=
	\int d^3 r
	\Psi^\dagger (\mathbf{r})
	\left(
	\frac{\left(\hat{\mathbf{p}} - \mathbf{A} \right)^2}{2m}
	-\frac{\kappa^2}{2m}
	+
	V(\mathbf{r})
	\right)
	\Psi (\mathbf{r})
	\notag \\
	&+
	\mathcal{H}_{\mathrm{int}},
\end{align}
where $\Psi (\mathbf{r}) =  (\psi_\uparrow (\mathbf{r}), \psi_\downarrow (\mathbf{r}))$ is the particle annihilation operator, $V(\mathbf{r})$ is the potential of the confining trap, and $\mathbf{A} = -\kappa (\sigma_x, \sigma_y, 0)$ is the non-Abelian vector potential, with coupling strength $\kappa$.
We assume s-wave interactions between particles,
\begin{align}
	\mathcal{H}_{\mathrm{int}}
	=
	\sum_{\sigma, \sigma^\prime = \uparrow, \downarrow}\frac{g_{\sigma \sigma^\prime}}{2} \int d^3 r \psi^\dagger_\sigma (\mathbf{r}) \psi^\dagger_{\sigma^\prime} (\mathbf{r}) \psi_{\sigma^\prime} (\mathbf{r}) \psi_\sigma (\mathbf{r}).
\end{align}

The Hamiltonian without interactions has a circle of degenerate single-particle ground states with momenta $\sqrt{p_x^2+p_y^2} = \kappa$ and $p_z = 0$.
The striped state, a superposition of two states with opposite momenta on the circle of degenerate single-particle ground states,  is the preferred ground state in the homogeneous ($V(\mathbf{r}) = 0$) infinite system,
if we take the renormalization of the interaction into account~\cite{Gopalakrishnan2011,Ozawa2011,Ozawa2012a}. The macroscopic wavefunction of the striped state with momenta $(\pm \kappa, 0, 0)$ is
\begin{align}
	\Psi_s (\mathbf{r})
	=
	\sqrt{n}
	\begin{pmatrix}
	\cos (\kappa x) \\ -i\sin (\kappa x)
	\end{pmatrix}, \label{stripedwave}
\end{align}
where $n$ is the density of particles.

\begin{figure}[htbp]
\begin{center}
\includegraphics[width=7.0cm]{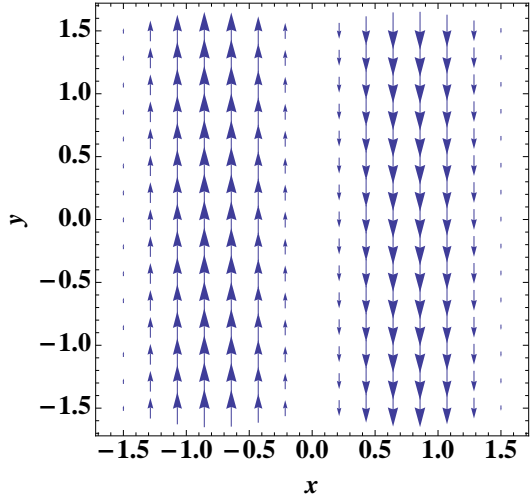}
\caption{The particle current $\mathbf{j}(\mathbf{r})$ of a striped state as a function of $x$ and $y$ in units of $\kappa^{-1}$.  The thickness of the arrows indicates the magnitude of the current.}  
\label{striped_flow}
\end{center}
\end{figure}

The particle current in the presence of the gauge field,
\begin{align}
	\mathbf{j}_{st}(\mathbf{r})
	&=
	\frac{1}{2im}(\Psi_s^* \nabla \Psi_s - \nabla \Psi_s^*  \,\Psi_s)
	-
	\frac{1}{m}\Psi_s^* \mathbf{A} \Psi_s
	\notag \\
	&=
	\left(
	0, -\frac{n\kappa}{m}\sin (2\kappa x), 0
	\right), \label{currentstriped}
\end{align}
does not vanish;
there is a local flow of particles in the striped state all the way out to $y=\pm\infty$. (We set $\hbar=1$ throughout.)
Figure \ref{striped_flow} plots $(j_x (\mathbf{r}), j_y (\mathbf{r}))$ as a function of $\kappa x$ and $\kappa y$.
The flow in the y direction alternates as we move along the x axis.
In a finite-size box, this flow would drive particles to the boundaries in the y direction, accumulating particles at the edge.  Thus the striped state {\it per se} is not a stable steady state in a finite-size system.
On the other hand, experimentally realizable ultracold atomic systems are inevitably finite,
and we are in need of finding a physically acceptable state which is similar to the striped state.

The particle circulation in the striped state is not quantized, since the local velocity, $\mathbf{v}_{st} \equiv \mathbf{j}_{st}/n$ is
\begin{align}
	\nabla \times \mathbf{v}_{st}
	=
	\left(0, 0, -\frac{2\kappa^2}{m} \cos (2\kappa x) \right).
	\label{curlvs}
\end{align}
Such nonquantized vorticity is a generic feature of systems coupled to external gauge fields, e.g., type-II superconductors.

Nonvanishing local flow is a common feature of striped states in general Rashba-Dresselhaus spin-orbit coupling fields of the form $\mathbf{A} = -\kappa (\sigma_x, \eta \sigma_y, 0)$, where $0 < \eta \le 1$.
When $\eta = 0$, i.e., an equal mixture of the Rashba and Dresselhaus spin-orbit couplings,
the local current vanishes.  In the experimentally realized spin-orbit coupling to date is an equal mixture of the Rashba and Dresselhaus spin-orbit couplings~\cite{Lin2011,Williams2012,Wang2012arXiv} (with no problem of particles accumulating at the boundaries).

\begin{figure}[htbp]
\begin{center}
\includegraphics[width=7.0cm]{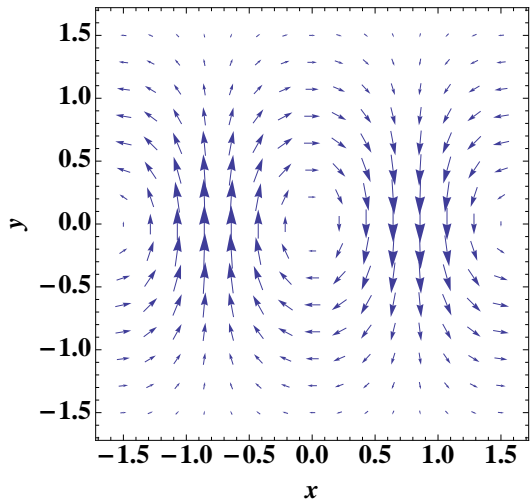}
\caption{The particle current $\mathbf{j}(\mathbf{r})$ of the modified striped state (\ref{mod_stripedwave}), with $\zeta = 1$, 
 as a function of $x$ and $y$ in units of $\kappa^{-1}$.  The magnitude of the current, shown by the thickness of the arrows, falls off at large $|y|$. Note the line of vortices along the x-axis.}
\label{mod_striped_flow}
\end{center}
\end{figure}

{\em Modified striped state.}
We now assume that the system is confined in the y direction by a harmonic potential, $V(\mathbf{r}) = m\omega^2 y^2/2$.  Taking the striped state (\ref{stripedwave}) as a starting point, we assume a macroscopic wavefunction for a modified striped state of the form
\begin{align}
	\Psi_{ms} (\mathbf{r})
	=
	\gamma (x,y) e^{i\phi (x,y)}
	\begin{pmatrix}
	\cos (\kappa x) \\ -i\sin (\kappa x)
	\end{pmatrix}, \label{mod_stripedwave}
\end{align}
where $\gamma (x,y)$ and $\phi (x,y)$ are real functions of $x$ and $y$.
Going beyond this form of ansatz, for example by modifying the spin part of the wavefunction, is beyond the scope of this paper.
In this state, the particle density is  $\gamma^2$, and the particle current is
\begin{align}
	\mathbf{j} (\mathbf{r})
	&=
	\frac{\gamma^2}{m}\left[
	\nabla \phi + \left(0, - \kappa \sin (2\kappa x), 0 \right)
	\right]
	\notag \\
	&=
	\gamma^2 \left( \frac{\nabla \phi}{m}  + \mathbf{v}_{st} (\mathbf{r}) \right).
\end{align}
Two physical conditions on $\Psi_{ms}$ are that the current vanishes at large $|y|$ and,
in equilibrium, $\nabla \cdot \mathbf{j} = 0$.
A natural choice for a system harmonically confined in the y direction is
\begin{align}
	\gamma = c\, e^{ -\zeta \kappa^2 y^2/2},
\end{align}
where $\zeta$ is a positive dimensionless variational parameter, and the normalization $c$ is determined by fixing the total number of particles $N$, 
\begin{align}
	c^2
	=
	\frac{N\kappa}{L^2}\sqrt{\frac{\zeta}{\pi}} = \sqrt2 \,\bar n_{ms}.
\end{align}
Here $L$ is the
linear size of the system in the x and z directions, and $\bar n_{ms} = \int dy \gamma^4/ \int dy \gamma^2$ is a mean particle density.
With this form for $\gamma$, the condition
$\nabla\cdot \mathbf{j} = 0$ is satisfied for
\begin{align}
	\phi= \frac{\zeta \kappa y}{\zeta + 2} \sin (2\kappa x).
\end{align}
Then,
\begin{align}
	\mathbf{j}(\mathbf{r})
	=
	\frac{\kappa}{m}\frac{2\gamma^2}{\zeta + 2}
	\left(
	\zeta \kappa y \cos (2\kappa x), - \sin (2\kappa x), 0
	\right).
\end{align}
In the limit $\zeta\to 0$, 
we recover the striped state (\ref{stripedwave}) and current (\ref{currentstriped}).
Figure \ref{mod_striped_flow} plots the current for $\zeta = 1$ as a function of $\kappa x$ and $\kappa y$.
When $\kappa y$ is small, the current behaves similarly to that in the striped state and flows in the $\pm y$ directions.
However, as $|y|$ grows, the current bends towards the $\pm x$ directions, and particles no longer accumulate at the boundaries in the y direction.
Near the center, this state in a finite system behaves similarly to a striped state.

The modified striped state has a line of vortices, along the z-axis, at $\mathbf{r} = (\nu\pi/2\kappa, 0, 0)$, where $\nu$ takes integer values.
While the particle velocity at the center of the vortices vanishes, the density remains nonzero; the vortices are coreless.
The curl of the local velocity in the modified striped phase, $\mathbf{v} = \mathbf{j}/\gamma^2$, is
simply that of the original striped phase, given by Eq.~(\ref{curlvs})~\footnote{The increase of the x-velocity with $y$ suggests that the wavefunction at the trap edges could be improved by allowing flow, as well as bending of the pseudospin, into the z direction (T.L-Ho, private communication).}.

{\em Energy.}
We turn now to examine when the modified striped state is energetically favorable at zero temperature. In the mean field, the energy per particle of the modified striped state is
\begin{align}
	\langle E \rangle_{ms}
	=
	\left\{
	\epsilon_k
	\frac{\zeta}{2}
	\left(
	1
	-
	\frac{1}{\zeta+2}
	\right)
	+
	\epsilon_i \sqrt{\zeta}
	\right\}
	+
	\frac{\epsilon_t}{\zeta}, \label{emstrap}
\end{align}
where $\epsilon_k \equiv \kappa^2/2m$, 
\begin{align}
	\epsilon_i &\equiv \frac{N \kappa}{16 \sqrt{2\pi}L^2}\left(3g_{\uparrow \uparrow} + 3g_{\downarrow \downarrow} + 2g_{\uparrow \downarrow} \right)
\end{align}
and $\epsilon_t \equiv m\omega^2/4\kappa^2$.

First we compare the kinetic and interaction terms in braces in Eq.~(\ref{emstrap}) with the corresponding terms of the simple striped state (\ref{stripedwave}), whose
 energy per particle is
\begin{align}
	\langle E \rangle_{s}
	=
	\frac{n}{16}\left( 3g_{\uparrow \uparrow} + 3g_{\downarrow \downarrow} + 2g_{\uparrow \downarrow}\right).
\end{align}
The kinetic energy of the modified striped state is positive and thus larger than that (zero) of the simple striped state.
Furthermore, the ratio of the interaction energy in the modified striped phase to that in the simple striped phase is $\bar n_{ms}/n$.
More relevant is to compare the energy of the modified state with a straightforward extension of the striped state with a {\it Gaussian-striped} state described by the condensate wavefunction,
\begin{align}
	\Psi_{gs} (\mathbf{r})
	=
	c\, e^{-\zeta \kappa^2 y^2/2}
	\begin{pmatrix}
	\cos (\kappa x) \\ -i\sin (\kappa x)
	\end{pmatrix}. \label{gswave}
\end{align}
Although this state is a natural extension of the striped state with a Gaussian decay, the divergence of the current does not vanish, and is not physically acceptable.
Nonetheless, since the correct ground state must have lower energy than the Gaussian-striped state,
we compare the energies of the Gaussian-striped and the modified striped states.
The energy of the Gaussian-striped state is
\begin{align}
	\langle E\rangle_{gs}
	&=
	\epsilon_k
	\frac{\zeta}{2}
	+
	\epsilon_i \sqrt{\zeta}
	+
	\frac{\epsilon_t}{\zeta}.
\end{align}
The energy difference between the modified striped state and the Gaussian-striped state arises just from the kinetic energies,
\begin{align}
	\langle E\rangle_{ms} - \langle E\rangle_{gs}
	=
	-\frac{\kappa^2}{2m}
	\frac{\zeta}{2(\zeta+2)} <0.
\end{align}
Thus, modifying the striped state with a spatially dependent phase to make the current divergence-free decreases the energy below that of the Gaussian extension of the striped state, showing that our ansatz wavefunction for the modified striped state is a promising weakly confined striped-like ground state.
 
We also compare the energy of the modified striped state with that of  a Gaussian extension of the plane-wave state, described by the wavefunction
\begin{align}
	\Psi_{gp}
	=
	c\, e^{-\zeta \kappa^2 y^2/2}
	\frac{e^{i\kappa x}}{\sqrt{2}}
	\begin{pmatrix}
	1 \\ -1
	\end{pmatrix}.
\end{align}
The current vanishes in this phase and it is thus physical.
Its energy is
\begin{align}
	\langle E\rangle_{gp}
	=
	\epsilon_k
	\frac{\zeta}{2}
	+
	\frac{N\kappa \sqrt{\zeta}}{8\sqrt{2\pi}L^2}
	\left(g_{\uparrow \uparrow} + g_{\downarrow \downarrow} + 2g_{\uparrow \downarrow} \right)
	+
	\frac{\epsilon_t}{\zeta}.
\end{align}
The kinetic energy and trap energy are the same as for the Gaussian-striped state.
The interaction energy of the plane-wave state is lower when $g_{\uparrow\downarrow} < (g_{\uparrow\uparrow} + g_{\downarrow\downarrow})/2$, in which case the mean-field ground state in an infinite system is also a plane-wave state~\cite{Wang2010}.
Thus, within mean-field coupling with weak confinement, the modified stripe phase is preferred over the Gaussian-plane-wave ground state if (and only if) the interspecies interaction is larger than the average intraspecies interaction.

\begin{figure}[htbp]
\begin{center}
\includegraphics[width=8.5cm]{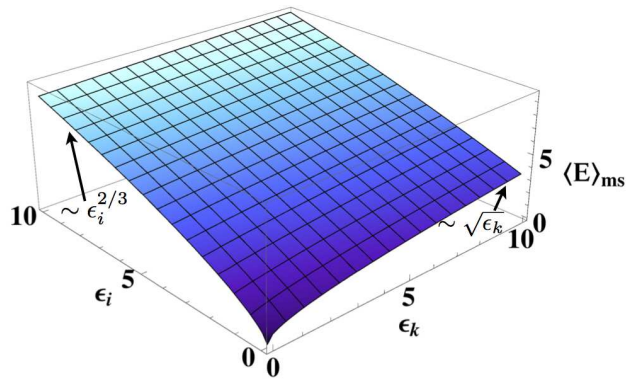}
\caption{The energy of the modified striped state $\langle E \rangle_{ms}$ as a function of the kinetic and interaction energies; all energies are in units of $\epsilon_t$.}
\label{emstrapg}
\end{center}
\end{figure}

\begin{figure}[htbp]
\begin{center}
\includegraphics[width=8.5cm]{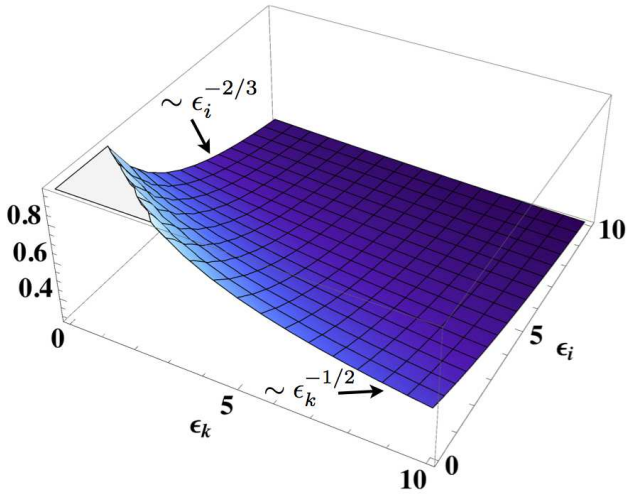}
\caption{The inverse width, $\zeta$, of the Gaussian $\gamma$, as a function of the kinetic and interaction energies in units of $\epsilon_t$; note that the figure is turned 90$^\circ$ from Figs.~\ref{emstrapg} and \ref{deltaeg}. The surface rises indefinitely as the kinetic and interaction energies become small.}
\label{zeta}
\end{center}
\end{figure}

Finally, we determine $\zeta$ variationally; the minimization of $\langle E\rangle_{ms}$ with respect to $\zeta$ implies that the optimal $\zeta$ is a function of the dimensionless ratios $\epsilon_k/\epsilon_t$ and $\epsilon_i/\epsilon_t$.  Figure~\ref{emstrapg}
shows the resulting energy of the modified striped state, $\langle E \rangle_{ms}/\epsilon_t$, as a function of $\epsilon_k/\epsilon_t$ and $\epsilon_i/\epsilon_t$, while Fig.~\ref{zeta} shows the optimal $\zeta$.

As $\omega \to 0$, the energy (\ref{emstrap}) is minimized when $\zeta = 0$, which is the simple striped state; as can be seen in Fig.~\ref{zeta},
$\zeta$ grows with increasing $\omega$.
In the limits of kinetic energy dominant over the interaction energy, and vice versa, we can obtain the asymptotic behaviors of $\zeta$ and the corresponding energy analytically.  

First, when $\epsilon_k\gg\epsilon_i$, by minimizing (\ref{emstrap}) and ignoring the interaction term, one obtains $\zeta \simeq 2 \sqrt{\epsilon_t / \epsilon_k} = \sqrt{2} m\omega / \kappa^2$. The energy to leading order is $\langle E \rangle_{ms} \simeq \omega / 2\sqrt{2}$.
This asymptotic behavior is valid when the estimated interaction energy using the asymptotic value of $\zeta$ is smaller than the kinetic and the trap energies, which yields the condition $\epsilon_i/\epsilon_t \ll (\epsilon_k/\epsilon_t)^{3/4}/2$. 
In the other limit, $\epsilon_i\gg\epsilon_k$, a similar analysis leads to $\zeta \simeq (2\epsilon_t /\epsilon_i)^{2/3}$ and $\langle E \rangle_{ms} \simeq 3(\epsilon_i^2 \epsilon_t/4)^{1/3}$.
The condition at which this asymptotic behavior is valid is $\epsilon_i/\epsilon_t \gg (\epsilon_k/\epsilon_t)^{3/4}/3$.
In this limit, the extent of the Gaussian in y is independent of $\kappa$.
For typical experimental parameters of spin-orbit coupled $^{87}$Rb~\cite{Williams2012},  $\omega/2\pi \sim 50$ Hz, $\kappa \sim \sqrt{2}\pi/800$nm, $N \sim 5\times 10^5$, and $L \sim 100\mu m$,
one has $\epsilon_k /\epsilon_t \sim 10000$ and $\epsilon_i / \epsilon_t \sim 2300$;  here one is in the latter limit, with $\zeta \sim 0.008$.

\begin{figure}[htbp]
\begin{center}
\includegraphics[width=8.5cm]{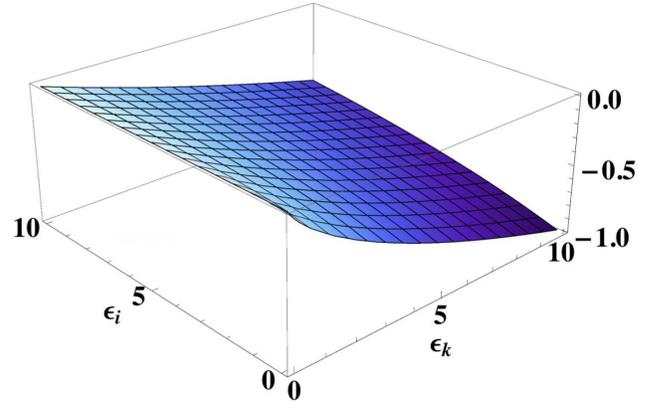}
\caption{The energy difference of the modified striped state and the Gaussian-striped state, $\langle {E}\rangle_{ms} - \langle {E}\rangle_{gs}$;  again all energies are in units of $\epsilon_t$.}
\label{deltaeg}
\end{center}
\end{figure}

The optimal value of $\zeta$ for the Gaussian-striped phase is similarly a function of the dimensionless ratios $\epsilon_k/\epsilon_t$ and $\epsilon_i/\epsilon_t$.  In Fig.~\ref{deltaeg}, we show the difference of the energy of the modified striped state and the Gaussian-striped state in a trap, as a function of $\epsilon_k/\epsilon_t$ and $\epsilon_i/\epsilon_t$.
The energy of the modified striped state always lies below that of the Gaussian-striped state, with the difference becoming larger as the kinetic energy becomes large compared to the trap energy.

In summary, the modified striped state is not only physically allowed, but it is also an energetically favorable starting description of steady-state striped states in a harmonic trap.
The arguments given in this paper hold in two as well as three dimensions, {\it mutatis mutandis}, and thus the state (\ref{mod_stripedwave}) is the preferred striped-like ground state in either situation.
As we noted earlier, the modified striped state has a line of coreless vortices, which can be probed to identify the state in possible future experiments.
 
\begin{acknowledgements}
We are grateful to Jason Ho for illuminating discussions, and to Xuebing Luo for pointing out a simple
error in Eq.~(12) of the published version of this paper.
This research was supported in part by NSF Grant No. PHY09-69790. 
\end{acknowledgements}

\end{document}